\documentclass[showpacs,preprintnumbers,amsmath,amssymb,prb,twocolumn]{revtex4}

\def\XXint#1#2#3{{\setbox0=\hbox{$#1{#2#3}{\int}$}
     \vcenter{\hbox{$#2#3$}}\kern-.5\wd0}}

\usepackage{graphicx}
\usepackage{subfigure}
\usepackage{bm}

\begin{document}

\title{Self-consistent T-matrix approach to gap renormalization in quantum magnets with bond disorder}

\author{O.\ I.\ Utesov$^{1,2}$}
\email{utiosov@gmail.com}

\affiliation{$^1$National Research Center ``Kurchatov Institute'' B.P.\ Konstantinov Petersburg Nuclear Physics Institute, 188300 Gatchina , Russia}
\affiliation{$^2$Department of Physics, St.~Petersburg State University, 7/9 Universitetskaya nab., St.~Petersburg 199034, Russia}

\date{\today}

\begin{abstract}

Based on the self-consistent T-matrix approximation (SCTMA), analytical theory of density of states (DOS) in three-dimensional quantum magnets with the bond disorder is proposed. It successfully describes DOS in both cases of resonant and non-resonant scattering which appearance is governed by the ratio of scattering length and the average distance between impurities. Corrections to the quasiparticles band gap in these cases are shown to be $\propto c^{2/3}$ and $\propto c$, respectively. Moreover, the theory yields a semi-circle form of DOS for the bound states inside the gap which results in highly nontrivial DOS in the intermediate parameters region between two limiting cases when the band DOS and the semi-circle are overlapped. Long-wavelength excitations are discussed. In the resonant regime their damping is almost constant $\propto c^{2/3}$, which according to Ioffe-Regel criterion means their localization. Applicability of the theory is illustrated by a quantitative description of the recent experimental data on spin-dimer system Ba$_{3-x}$Sr$_x$Cr$_2$O$_8$.

\end{abstract}

\maketitle

\section{Introduction}
\label{SIntro}

Quantum magnets are among most extensively studied both theoretically and experimentally compounds of the last several decades (see Ref.~\cite{vasiliev2018} and references therein). In particular, they allow to investigate quantum phase transitions~\cite{sachdev2011}, established themselves as very convenient for such purposes systems~\cite{giamarchi2008bose}. Moreover, disorder in magnetic insulators can be introduced in more or less controllable way, which allows studying so-called ``dirty-boson'' physics(see Ref.~\cite{zheludev2013dirty} and references therein). Notably, in bromine doped Dichlorotetrakis-thiourea nickel (DTN) such elusive phase as Bose-glass~\cite{fisher1989} was observed experimentally~\cite{yu2012bose}. Moreover, it was proven that in systems with quenched disorder glassy phase should always intervene in the direct Mott insulator - superfluid transition~\cite{theorem}. The problem of phase transitions to glassy phases is scrutinized theoretically in many papers (see, e.g., Refs.~\cite{fisher1989,kruger2011,thomson2015,syromyatnikov2017,syromyatnikov2017Mott,yashenkin2016}).

Bond disorder in magnetic insulators can be introduced by replacing some atoms involved into superexchange paths~\cite{hong2010,yu2012bose, huvonen2012, grundmann2013}. It results in change of corresponding magnetic interaction constants~\cite{povarov2015dynamics}. The advantage of this method is a possibility to control impurities concentration in a specimen, which is important for studying peculiar predictions of dirty-boson physics~\cite{zheludev2013dirty}. For instance, as it was shown in Ref.~\cite{Povarov2017}, in Br doped DTN with high concentration of dopants ($\approx 21 \%$) gapped Mott insulator phase vanishes. Importantly, bond disorder prepared following the described method should be a kind of the binary one rather than continuously distributed.

Recently a theoretical description of the binary disorder influence on the elementary excitations in quantum magnets was proposed~\cite{utesov2014}. It was based on conventional T-matrix method~\cite{izyumov,doniach1998} and for 3D systems yields correction proportional to impurity concentration $c$ for the system gap and the damping $\propto c$ and momentum distance to the spectrum minimum $\mathbf{k}_0$. This theory can be used for experimental data description. However, those result are inapplicable in the range of parameters providing so-called resonance scattering (in the vicinity of the threshold of bound  state on impurity appearance).

In the present paper we propose a theory which is capable to describe both resonant and non-resonant scattering regimes. It is based on the self-consistent T-matrix approximation (SCTMA) (see, e.g., Refs.~\cite{lee1993,ostrovsky2006,yashenkin2016}) which establish itself as a powerful method, although, its results can be usually a subject of further corrections~\cite{yashenkin2001}.

We observe that the system density of states (DOS) and, correspondingly, the gap behaviour is drastically different with predictions of standard T-matrix approach~\cite{utesov2014} when the scattering on impurities is close to the resonant one. Moreover, we show that the results are governed by the ratio of scattering length and mean distance between defects $\propto c^{-1/3}$. When this ratio is large correction to the system gap is $\propto c^{2/3}$ while in the opposite case conventional $\propto c$ behaviour restored. For large enough disorder strengths SCTMA also yields semi-circle law for bound state on impurity DOS with broadening being $\propto c^{1/2}$ (c.f. with the theory of electrons on Landau levels broadening~\cite{ando1974}). In the intermediate parameters region the semi-circle and band DOS start to overlap which results in highly non-trivial DOS. Next, we discuss long-wavelength elementary excitations. We show that in the resonant regime broadening is almost constant $\gamma_\mathbf{k} \propto c^{2/3} + O(k^2)$ for excitations with momenta $ k \ll c^{1/3}$. It is shown that according to Ioffe-Regel criterion~\cite{ioffe1960} it means their localization. While in the crossover regime linewidths of long-wavelength excitations remain smaller but finite, in the non-resonant one we obtain previous result~\cite{utesov2014} $\gamma_\mathbf{k} \propto c k$. Finally, we successfully describe experimental findings of Ref.~\cite{gazizulina2017} for gap renormalization in spin-dimer system Ba$_{3-x}$Sr$_x$Cr$_2$O$_8$.

The rest of the paper is organized as follows. In Section~\ref{SBas} we briefly discuss spin-dimer systems with bond disorder and its treatment using T-matrix method. Section~\ref{STheor} is devoted to the self-consistent T-matrix approach technique for system density of states calculations in quantum magnets with diagonal disorder. We introduce important spatial scales of the problem and present simple results in limiting cases. We also discuss line-shape and broadening of elementary excitations. In Section~\ref{SAppl} we apply the developed theory to the experimental data for Ba$_{3-x}$Sr$_x$Cr$_2$O$_8$. Section~\ref{SSum} contains summary of the results. Cumbersome solutions of the cubic SCTMA equation can be found in Appendix.

\section{Basic formalism}
\label{SBas}

\subsection{Triplon spectra and bond disorder in spin-dimer systems}

As an example of systems for which our analysis is applicable we consider spin-dimer systems. In this subsection we briefly remind their basic properties and introduce disorder.

Heisenberg Hamiltonian of spin-dimer systems has the standard form~\cite{giamarchi2008bose}:
\begin{equation}  \label{ham1}
  \mathcal{H}_0=\sum_i J_0 {\mathbf S}_{i,1} \cdot \mathbf{S}_{i,2} + \sum_{\langle i,j\rangle}  J_{ij} \left( \mathbf{S}_{i,1} \cdot \mathbf{S}_{j,1} + \mathbf{S}_{i,2} \cdot \mathbf{S}_{j,2}\right),
\end{equation}
where $i$ and $j$ are the neighboring dimers, $J_0$ and $J_{ij}$ are intra- and inter- dimer exchange couplings, respectively, usually $J_0\gg |J_{ij}|$ is assumed. Moreover, we write the interdimer interaction in the simplest form because it does not affect the results below (see, however, Ref.~\cite{utesov2014jmmm}). We also do not consider external magnetic field $H$  which effect in the gapped phase is the simple shift of triplon spectra $\pm g \mu_B H$ (see, e.g., Ref.~\cite{giamarchi2008bose} and references therein).

One can introduce three pairs of bosonic operators describing creation and annihilation of the triplons~\cite{sachdev1990,kotov1998}, $\mathfrak {a}|0\rangle=\mathfrak{b}|0\rangle=\mathfrak{c}|0\rangle=0$, $\mathfrak {a}^+|0\rangle=|\uparrow\uparrow \rangle$, $\mathfrak{b}^+|0\rangle=| \downarrow\downarrow \rangle$, and $\mathfrak {c}^+|0\rangle=\frac{1}{\sqrt{2}}\left(|\uparrow\downarrow \rangle+|\downarrow\uparrow \rangle \right)$. At zero external magnetic fields the spectrum of the triplons is threefold degenerate. In the calculations in Sec.~\ref{SAppl} we shall use the RPA one~\cite{kofu2009},
\begin{equation}\label{spec1}
  \varepsilon_\mathbf{k} = \sqrt{J^2_0 + J_0 J_\mathbf{k}}.
\end{equation}
Here $J_\mathbf{k}$ is Fourier transform of interdimer interactions, and momenta $\mathbf{k}$ are taken dimensionless (lattice parameters are put to be equal to one). So, minimum point of this spectrum in spin-dimer compounds usually reads $\mathbf{k}_0 = (\pi,\pi,\pi)$.

Next, lets introduce binary bond disorder to the system above. In particular, we will treat the system with some amount of different intradimer couplings measured by their dimensionless concentration $c$. So, the Hamiltonian now reads
\begin{equation}
  \mathcal{H}=\mathcal{H}_0+V,
  \label{impHam}
\end{equation}
where perturbation $V$ is due to the disorder and
\begin{equation}	\label{pert1}
  V = \sum_{\{n\}} u \mathbf{S}_{n,1} \cdot \mathbf{S}_{n,2}.
\end{equation}
Here $u$ measures deviation of $J_0$ on imperfect bonds, which are denoted by $\{n\}$. In the bosonic form~\cite{utesov2014}
\begin{equation} \label{pert2}
  V = u \sum_{\{n\}}\mathfrak{a}^+_n \mathfrak{a}_n,
\end{equation}
and the same contributions for other triplon branches. We notice that in the linear spin waves theory triplons interaction is negligible, so we can treat one certain triplon branch.

Diagonal type of disorder~\eqref{pert2} can be also introduced to other quantum magnets, e.g., antiferromagnets with large single-ion easy-plane anisotropy, see Ref.~\cite{utesov2014}.

\subsection{T-matrix}

At small impurities concentration $c \ll 1$ standard T-matrix method~\cite{izyumov,doniach1998} can be utilized for calculations of triplon spectrum corrections due to the scattering on defects (see Ref.~\cite{utesov2014} for the details). Here we briefly remind basic ideas.

T-matrix approach is based on exact solution of one impurity problem which can be simply expressed via quasiparticle Green's function
\begin{equation}\label{green2}
  G^{-1}(\omega, \mathbf{k}) = \omega - \varepsilon_\mathbf{k} - T(\omega, \mathbf{k}),
\end{equation}
where in case of disorder only in intradimer couplings averaged over disorder configurations self-energy $T(\omega)$ reads
\begin{equation}
  T(\omega)=\frac{c u}{1- u \int \frac{d^3 q}{(2 \pi)^3} G_0(\omega, \mathbf{q})}.
  \label{T1}
\end{equation}
Here $G_0(\omega, \mathbf{k})$ is bare Green's function,
\begin{equation}\label{green1}
  G_0(\omega, \mathbf{k}) = \left( \omega - \varepsilon_\mathbf{k} - i 0 \right)^{-1}.
\end{equation}
Then, corrections to quasiparticles energy and damping have the form:
\begin{equation} \label{encor}
   E_\mathbf{k}=\varepsilon_{\mathbf{k}}+ \mathrm{Re} \, T(\varepsilon_{\mathbf{k}}),
  \quad
  \gamma_\mathbf{k}= \mathrm{Im} \, T(\varepsilon_{\mathbf{k}}).
\end{equation}
In 3D case the integral in Eq.~\eqref{T1} is convergent for every $\omega$. In the vicinity of the spectrum~\eqref{spec1} minimum it gives finite correction to the gap value and linear in momentum damping of triplons~\cite{utesov2014}.

In order to have some preliminary insights for subsequent SCTMA calculations it is instructive to perform simple analysis for ``Debye-type'' model of triplon spectra with spherical Brillouin zone characterized by a single parameter $k_D$ of the order of unity. Simplified dispersion near its minimum reads
\begin{equation}\label{specD}
  \varepsilon_\mathbf{k} = \Delta_0 + \alpha k^2,
\end{equation}
where $\mathbf{k}$ is a deviation of momentum from $\mathbf{k}_0$. Then, one can easily calculate the integral in Eq.~\eqref{T1} and obtain
\begin{eqnarray}
  &-& \frac{1}{2 \pi^2 \alpha}\left[ k_D - \int^{k_D}_0 dq \frac{(\Delta_0 - \omega)/\alpha}{q^2 + (\Delta_0 - \omega)/\alpha} \right] \approx \nonumber \\
  &-& \frac{1}{4 \pi \alpha}\left[ k^\prime_D - \sqrt{\frac{\Delta_0 - \omega}{\alpha}}\right],  \label{int1}
\end{eqnarray}
provided that condition $|\Delta_0 - \omega|/\alpha \ll 1$ is satisfied. We also introduce $k^\prime_D = k_D/ (\pi/2) \sim 1$. Notice that for any realistic model approximation~\eqref{specD} essentially fails near the Brillouin zone boundary and the constant $k^\prime_D$ from the upper limit of integration should not be treated very seriously, although it can be calculated numerically for each particular gapped spectrum~\eqref{spec1}.

Next, using Eq.~\eqref{int1} we can derive spectrum correction~\eqref{encor}. Renormalized gap and damping are given by (cf. Ref.~\cite{utesov2014})
\begin{equation} \label{encor2}
   \Delta = \Delta_0 + \frac{c u}{\displaystyle{1 + \frac{u}{4 \pi \alpha} k^\prime_D}},
  \quad
  \gamma_\mathbf{k}=\frac{ \displaystyle{\frac{c u^2}{4 \pi \alpha}} }{\left(\displaystyle{1 + \frac{u }{4 \pi \alpha}k^\prime_D} \right)^2} k.
\end{equation}
However, if the denominator in the these formulas is small, correction to the gap value and quasiparticle damping become very large and the conventional T-matrix approach is inapplicable. This is a manifest of a resonant scattering regime, which can be observed when the parameter
\begin{equation}\label{b1}
  b \equiv k^\prime_D + \frac{4 \pi \alpha}{u}
\end{equation}
is close to zero. Importantly, parameter $b$ provides spatial scale for the theory --- scattering length equal to $1/b$, which becomes very large in the resonant regime and is of the order of lattice parameter in the non-resonant one. Moreover, using this quantity it is easy to analyze a possibility of bound states on impurities appearance. The bound state energy is given by the pole of the T-matrix, which yields
\begin{equation}\label{bound1}
  \omega_{loc} = \Delta_0 - \alpha b^2
\end{equation}
for $k^\prime_D \gg b > 0$. It corresponds to negative $u$ satisfying condition $|u| > 4 \pi \alpha/ k^\prime_D$, positive $u$ being unable to produce the bound state with energy inside the gap.

In order to describe system properties in the whole range of parameters including resonant scattering regime we propose a theory based on the self-consistent T-matrix approximation (SCTMA). If one puts $G(\omega, \mathbf{q})$ instead of $G_0(\omega, \mathbf{q})$ in the integral in Eq.~\eqref{T1}, then it becomes an equation for $T(\omega)$. SCTMA was successfully used in gapped magnets in Ref.~\cite{yashenkin2016} for determining the correlation length (or gap value) of disordered system in the external magnetic field and for description of the transition to the Bose-glass phase.

\section{Self-consistent T-matrix approximation}
\label{STheor}

In the present case of diagonal disorder general equation of self-consistency reads
\begin{equation}
  T(\omega)=\frac{c u}{1 - u \int \frac{d^3 q}{(2 \pi)^3} [\omega - \varepsilon_\mathbf{k} - T(\omega)]^{-1}}.
  \label{ST1}
\end{equation}
Using similar to~\eqref{int1} trick it can be rewritten as
\begin{equation}
  T(\omega)=\frac{c u}{1 + \frac{u}{4 \pi \alpha} \left[ k^\prime_D - \sqrt{\frac{\Delta_0-\omega + T(\omega)}{\alpha}} \right]}.
  \label{ST2}
\end{equation}
Notice, that quasiparticles density of states (DOS) is proportional to integral in Eq.~\eqref{ST1}. So, it reads
\begin{equation}\label{gw1}
  \rho(\omega) = \frac{1}{4 \pi^2 \alpha} \mathrm{Im} \, \sqrt{\frac{\Delta_0-\omega + T(\omega)}{\alpha}}.
\end{equation}
Importantly, the condition for DOS to be nonzero coincides with $\mathrm{Im} \, T(\omega) \neq 0$ (see Eq.~\eqref{ST2}).

In order to simplify notation we introduce new variable $x = (\Delta_0-\omega)/\alpha$, dimensionless T-matrix $t(x) \equiv T(x)/\alpha$, impurities concentration-dependent quantity $\tilde{c} = 4 \pi c$, and we use the parameter $b$ defined in Eq.~\eqref{b1}. Then, after simple transformations one can write cubic equation for $t$:
\begin{equation}\label{STeq1}
  t^3 + t^2 (x-b^2) + 2 t b \tilde{c} - \tilde{c}^2 = 0.
\end{equation}
This equation can be solved analytically using Cardano's formula (see Appendix for the details), however corresponding solutions are cumbersome, and their physical meaning is obscure. So, we start our analysis with several limiting cases and then discuss intermediate parameters region. Importantly, limiting cases correspond to small and big ratios between scattering length $1/b$ and mean distance between impurities $c^{-1/3}$.

\subsection{Density of states}

\subsubsection{$b=0$}

\begin{figure}[t]
  \centering
  \includegraphics[width=8cm]{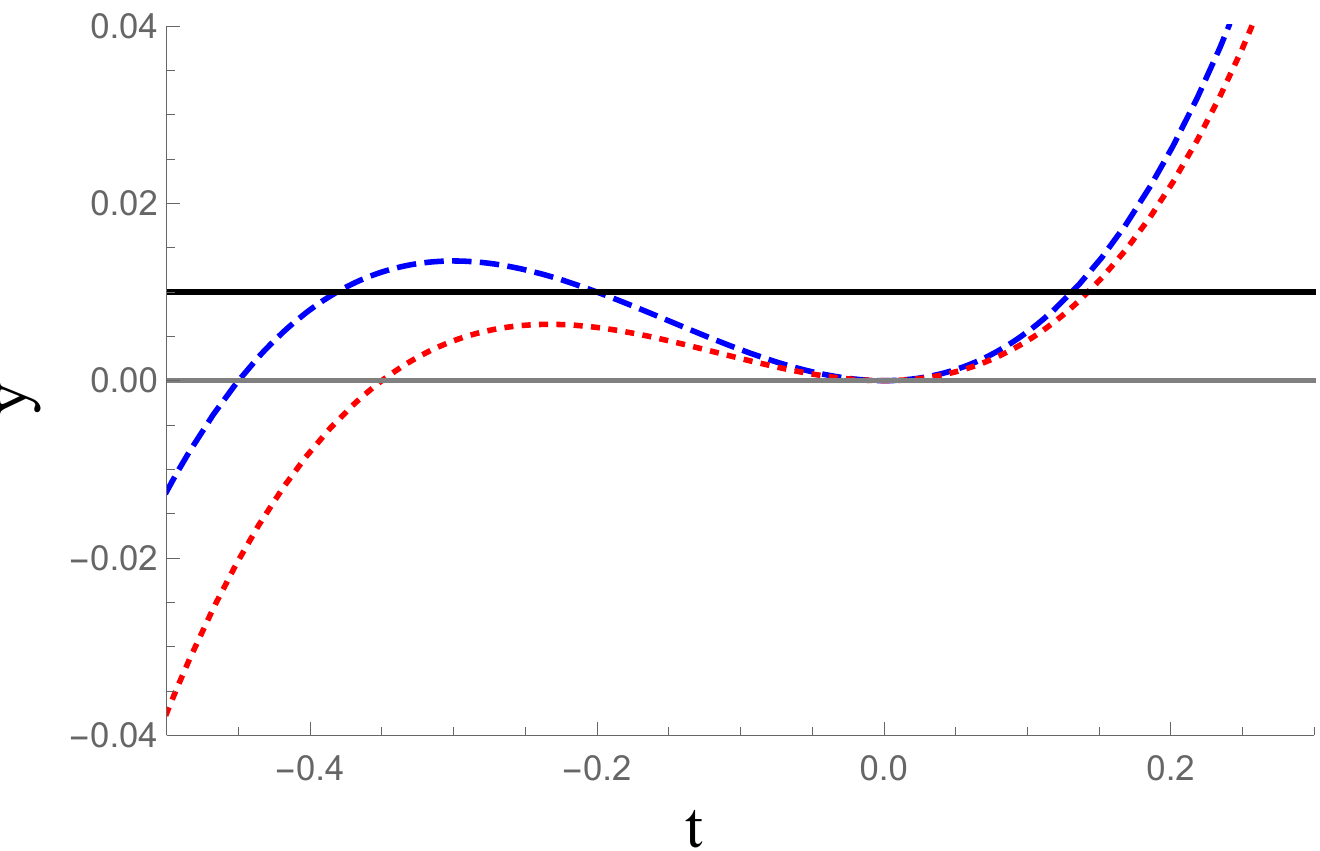}\\
  \caption{Graphic solution of Eq.~\eqref{STeq1} for $b=0$ and $\tilde{c}=0.1$. Solid black line is $y = \tilde{c}^2$, dashed blue curve stands for $f(t)$ (see Eq.~\eqref{ft1}) at $x=0.45$, dotted red one is the same for $x=0.35$, gray line is a guide for eyes. Evidently smaller $x$ yield two complex conjugate solutions $t$ which provide nonzero density of states at the corresponding frequencies.}\label{fig1}
\end{figure}

\begin{figure}[t]
  \centering
  \includegraphics[width=8cm]{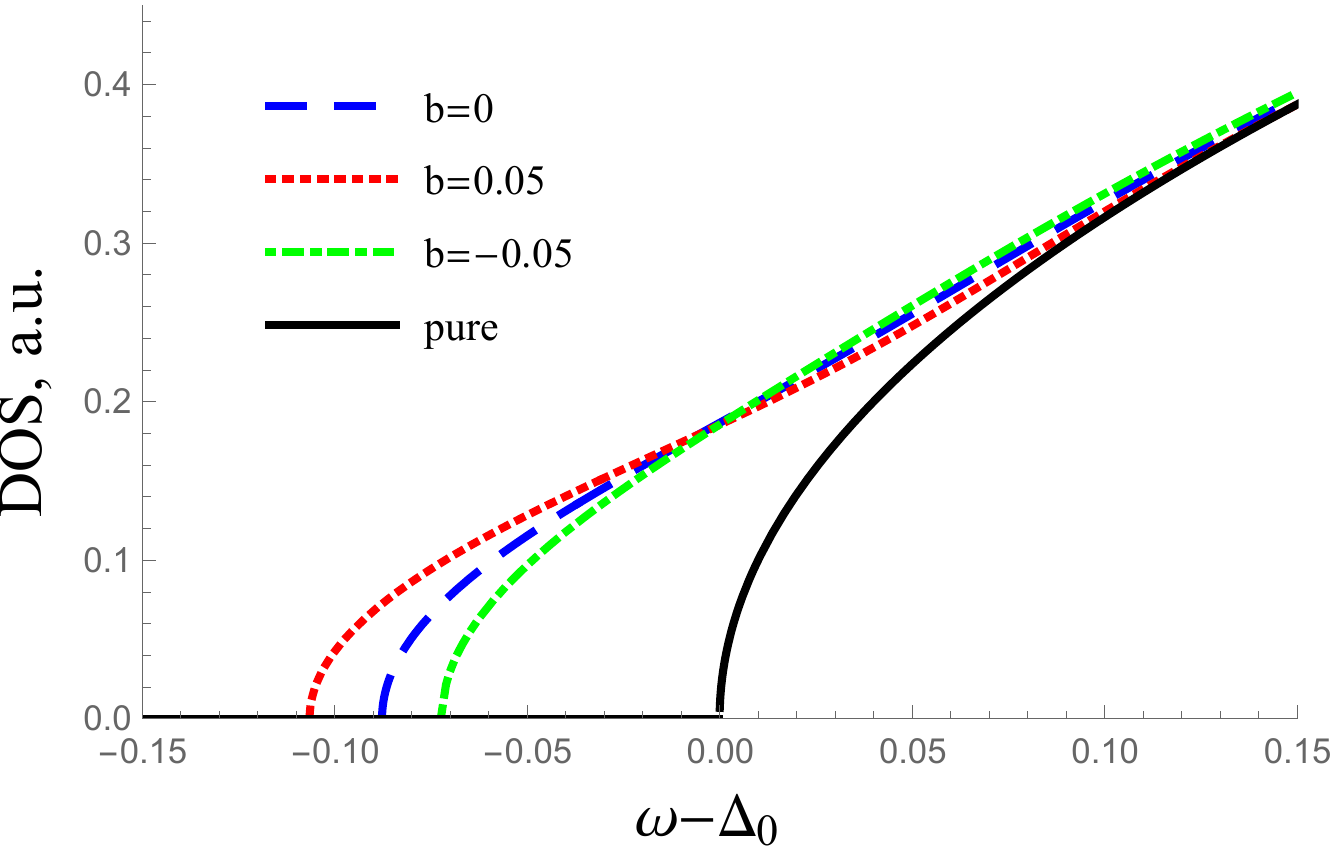}\\
  \caption{Density of states of the system (we take $\alpha=1, \tilde{c}=0.01$) at different $b$ values corresponding to the resonant scattering regime $|b| \ll \tilde{c}^{1/3} \approx 0.22$. Solid black line illustrates pure system DOS.}\label{fig2}
\end{figure}

If $b=0$ (exact resonance condition, see Eq.~\eqref{encor2}), then the analysis of Eq.~\eqref{STeq1} significantly simplifies. It can be performed graphically. We introduce the function with very simple behaviour
\begin{equation}\label{ft1}
  f(t) = t^3 + x t^2,
\end{equation}
real solutions are given by the cut of its plot with $y=\tilde{c}^2$ line, see Fig.~\ref{fig1}. Evidently for $x \leq 0$ there is always one real solution and two complex conjugate, with nonzero imaginary part. One of the latter is physical providing nonzero density of states in this region, namely, $\omega > \Delta_0$. In contrast, for large enough $x>0$ solutions with nonzero imaginary part disappear. This phenomenon is governed by the local maximum in $f(t)$ at $t= - 2 x/3$, so when
\begin{equation}\label{b0cond}
  f(-2x/3)= \frac{4}{27} x^3 > \tilde{c}^2 \Leftrightarrow x > 3 \left( \frac{\tilde{c}}{2} \right)^{2/3}
\end{equation}
all three solutions are real and $\rho(x)=0$. This condition defines renormalized gap value which reads
\begin{equation}\label{gapb0}
  x_0 = 3 \left( \frac{\tilde{c}}{2} \right)^{2/3} \Leftrightarrow \Delta = \Delta_0 - 3 \alpha  \left( 2 \pi c \right)^{2/3}.
\end{equation}
We report here unusual gap dependence on the impurity concentration, which is the result of the self-consistency. Next, using solutions presented in Appendix one can show that close to the gap ($x = x_0 - \delta x$ with $\delta x \ll x_0$) DOS is proportional to $\sqrt{\delta x} \propto \sqrt{\omega - \Delta}$ as in pure compound, see  dashed blue curve in Fig.~\ref{fig2}.

\subsubsection{$|b| \ll \tilde{c}^{1/3}$}

Based on the solution for $b=0$ one can show that the resonant regime governs the DOS behaviour at $|b| \ll \tilde{c}^{1/3}$. At frequencies near the gap, which we are mostly interested in, $t$ and $x$ are of the order of $\tilde{c}^{2/3}$. Thus, terms with $b$ in Eq.~\eqref{STeq1} can be treated as small perturbations. It is convenient to introduce $\tilde{x} = x - b^2$ and denote
\begin{equation}\label{ft1}
  f(t) = t^3 + \tilde{x} t^2 +2 t b \tilde{c},
\end{equation}
the last term being small correction. Once again analyzing the local maximum of this function and comparing it with $\tilde{c}^2$ one obtains perturbative correction to $x_0$ of Eq.~\eqref{gapb0},
\begin{equation}\label{gapbC}
  x_0 = 3 \left( \frac{\tilde{c}}{2} \right)^{2/3} + 2^{2/3}  b \tilde{c}^{1/3},
\end{equation}
and thus also small correction to the gap value~\eqref{gapb0}. DOS in this case is shown for $\tilde{c}=0.01, b=\pm 0.05$ in Fig.~\ref{fig2}, it has the same form as the one for $b=0$. We notice, that the gap values corresponding to curves shown in this figure for disordered system are well described by Eq.~\eqref{gapbC}.

Importantly, even for $b>0$ in this regime separated bound states with frequency~\eqref{bound1} do not appear. The physical reason is that scattering length in this case is much larger than the mean distance between defects, and possible in one impurity problem bound states are transformed into states extended over large volume.

\subsubsection{$|b| \gg \tilde{c}^{1/3}$}

\begin{figure}
  \centering
  \includegraphics[width=8cm]{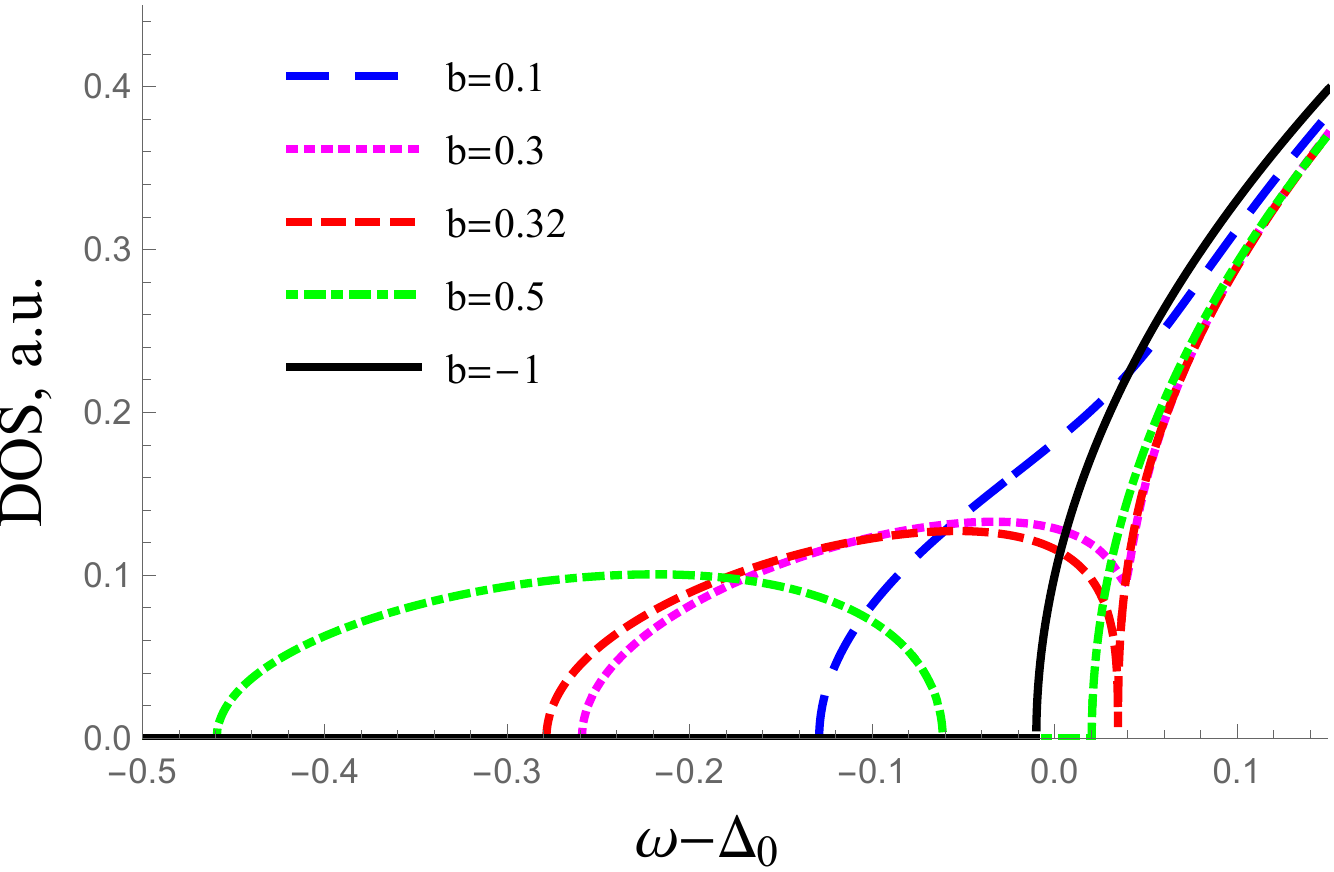}\\
  \caption{ Different ratios of $b$ and $\tilde{c}^{1/3}$parameters (or, in other words, scattering length $1/b$ and average distance between impurities $c^{-1/3}$) lead to various form of the system DOS. Large negative $b$ values (small negative disorder strengths) give only small correction to the band gap (solid black curve). When $b$ changes from negative to positive values throughout resonance region ($|b| \ll \tilde{c}^{1/3} $) DOS develops a feature connected with two parts of the spectrum - bound on impurities states and quasiparticles in the band. This peculiarity is most pronounced when $b \lesssim 3 \tilde{c}^{1/3}/2 \approx 0.32$ (dotted magenta and short-dashed red curve) in full agreement with the analytic solution. At larger $b$ bound states DOS is separated from the band one (dot-dashed green curve). Parameters $\alpha=1, \tilde{c}=0.01$ were used.
}\label{fig3}
\end{figure}

Here we will consider $b$ as a big parameter for Eq.~\eqref{STeq1}, which can be rewritten in the form
\begin{equation}\label{STeq2}
  \frac{t^3 + t^2 x}{b^2} = \left(t-\frac{\tilde{c}}{b}\right)^2.
\end{equation}
In this form it can be treated perturbatively using $t = \tilde{c}/b + \delta t$ ansatz. This yields
\begin{equation}\label{tb1}
  t \approx \frac{\tilde{c}}{b} \left( 1 + \frac{1}{b}\sqrt{\frac{\tilde{c}}{b} + x}\right).
\end{equation}
It results in the following gap renormalization:
\begin{equation}\label{gapbB}
  \Delta = \Delta_0 + 4 \pi \alpha \frac{c}{b},
\end{equation}
which is evidently the same with the one obtained within conventional T-matrix approximation, see Eq.~\eqref{encor2}. DOS is proportional to $\sqrt{\omega-\Delta}$, see  solid black curve in Fig.~\ref{fig3}). The latter is drawn for $\alpha=1, \tilde{c}=0.01, b=-1$; gap value is given with high accuracy by Eq.~\eqref{gapbB}.

However, if $k^\prime_D \gg b \gg \tilde{c}^{1/3}$ (it can be satisfied only in case of large enough negative $u$ being able to provide bound states inside the gap) other type of the solution exists. In Eq.~\eqref{STeq1} we can introduce small parameter $\delta x = x - b^2$, which describes how close is the frequency to the energy of localized on impurity state. It can be shown that in this region if $\mathrm{Im} \,t \neq 0 $ then $|t|, x \sim \sqrt{\tilde{c}}$, and $\tilde{c}^2$ in~\eqref{STeq1} is negligible. So, its solution is straightforward, and
\begin{equation}\label{tb2}
  t = \frac{\sqrt{\delta x^2 -8 b \tilde{c}} - \delta x }{2}.
\end{equation}
It gives famous (see, e.g., the theory of electron lines broadening on Landau levels~\cite{ando1974}) semi-circle form of the isolated level spectral weight with the center at $\omega_{loc}$ (see Eq.~\eqref{bound1}) and ``radius'' $\Delta \omega = \alpha \sqrt{32 \pi b c}$ which is much smaller than the distance to the quasiparticle band $\Delta - \omega_{loc} \approx \alpha b^2$. Explicitly,
\begin{equation}\label{DOSb1}
  \rho(\omega) \propto \sqrt{32 \pi b c - \left(b^2-\frac{\Delta_0-\omega}{\alpha}\right)^2}.
\end{equation}
This semi-circle law for DOS is a sign that the isolated level is broadened independently from other levels in continuum. However, in the present case in spite of $u<0$ the correction to band gap $\Delta_0$ is positive due to a repulsion between this two parts of the spectrum, see dot-dashed green curve in Fig.~\ref{fig3}. We point out, that the band gap in this case can be calculated using Eq.~\eqref{gapbB}.

\subsubsection{$ b \sim \tilde{c}^{1/3}$}

It is seen from Eqs.~\eqref{bound1},~\eqref{gapbB} and~\eqref{tb2} that if one starts changing $b$ from $b \gg \tilde{c}^{1/3}$ limit towards $b \sim \tilde{c}^{1/3}$ two parts of spectrum, namely, the band and bound on impurities states, become close to each other, and some kind of ``interaction'' between them appears. It results in the deformation of semi-circle as it is shown in Fig.~\ref{fig3} by dot-dashed green curve.  At further lowering of $b$ densities of states for isolated level and the band are essentially overlapped, which makes system DOS  much more complicated (see Fig.~\ref{fig3}, short-dashed red, dotted magenta, and long-dashed blue curves). 

The boundary between regimes of single overlapped DOS and the one with separated semi-circle can be found analytically. Corresponding cumbersome equations are presented in Appendix, they yield $b = 3 \tilde{c}^{1/3}/2$ curve dividing the regimes in the parameters space $(\tilde{c},b)$. However, for $b \gtrsim  3 \tilde{c}^{1/3}/2 $ we report that the band gap value can still be estimated using Eq.~\eqref{gapbB} and the lower boundary of semi-circle is given by $\omega_{loc} - \Delta \omega$ with high accuracy.

At the end of this Subsec. we would like to point out that in real systems DOS peculiarities discussed here will be smeared out due to other contributions to the self-energy. First, there is quasiparticles interaction, however it is usually small and it can be taken into account via spectrum~\eqref{specD} renormalization. Second, there are other contributions to the self-energy from the scattering on impurities which usually result in exponential tails in DOS (see, e.g., Refs.~\cite{zittartz1966,halperin1966,halperin1967, lifshitz1968, yashenkin2001, yashenkin2016}). Thus, an important question for further studies arises, to what extent dip in the DOS (visible e.g. for $b=0.3$ and $0.32$ in Fig.~\ref{fig3}) can be pronounced in real systems?

\subsection{Long-wavelength elementary excitations}

\begin{figure}
  \centering
  \includegraphics[width=8cm]{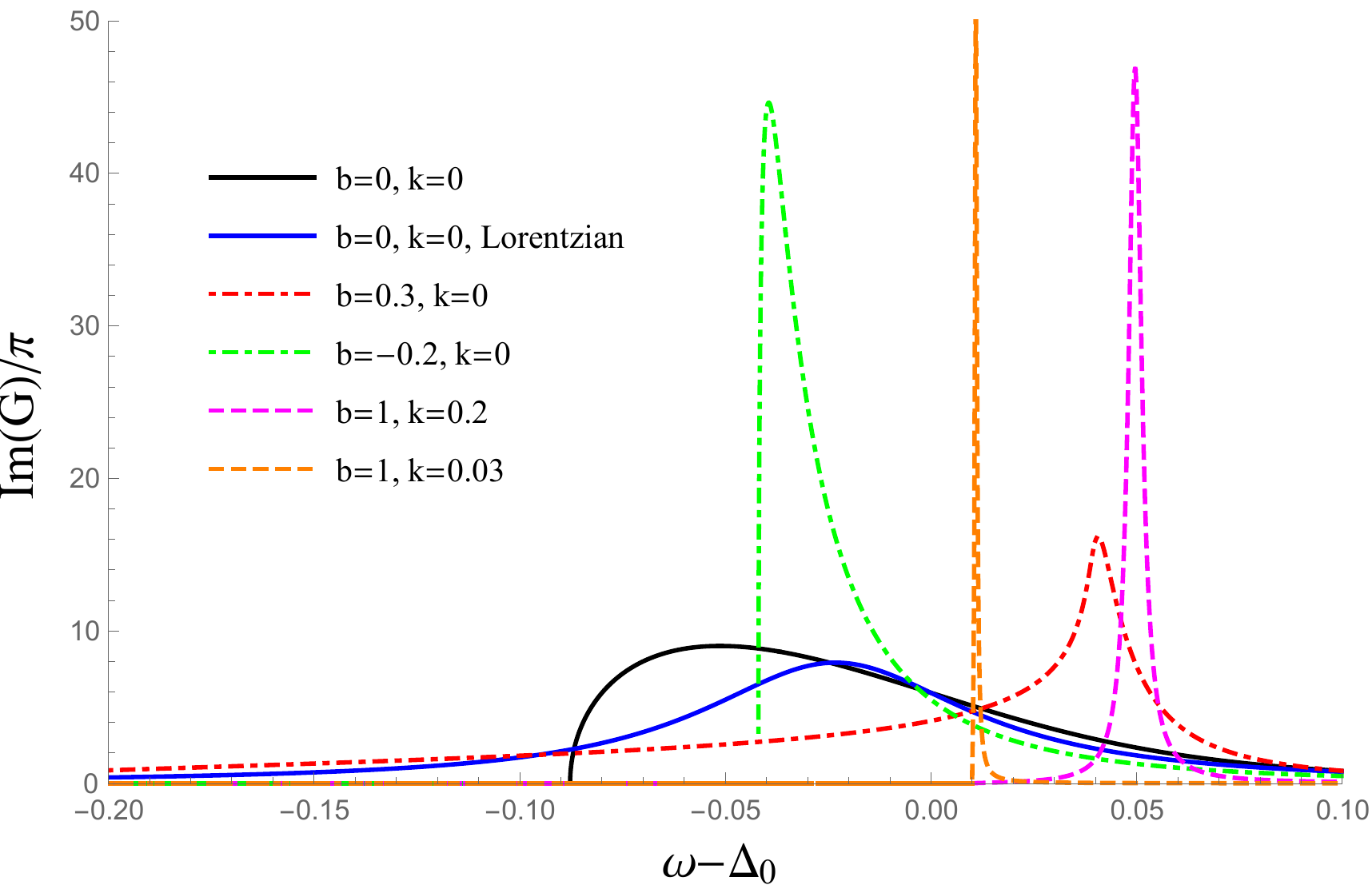}\\
  \caption{ Spectral weights for different parameters $b$ and momenta $k$. Solid black line shows spectral weight at $b=0, k=0$, which has unusual finite width. Imaginary part of $G(\omega,0)$ in the Lorentzian (on-shell) approximation (solid blue curve, plotted using Eq.~\eqref{green4}) describes well the line broadening in this case. In the crossover region $|b| \sim \tilde{c}^{1/3} \approx 0.22$ we plot spectral weight for $k=0$, and two values $b=0.3$ (dot-dashed red curve) and $b=-0.2$ (dot-dashed green curve), both also have finite linewidth. However, rather sharp peaks are formed, despite for $b=0.3$ one can see very wide ``shoulder'' as a consequence of complicated DOS (see Fig.~\ref{fig3}). For $b = 1 \gg \tilde{c}^{1/3}$ (non-resonant regime) we have well-defined peaks with small broadenings $\propto k$ (dashed orange and magenta curves, with manually decreased heights for illustration purposes). One can see significant asymmetry of the spectral weights (see text), except for $b=1, k=0.2$ curve, which is well separated from the gap. Parameters $\alpha=1, \tilde{c}=0.01$ were used.
}\label{fig4}
\end{figure}

After obtaining solutions of SCTMA equation in different regimes we can analyze spectral weights and broadenings corresponding to modes with small $k \ll k_D$ --- long-wavelength triplons.

One can rewrite Green's function $G(\omega, \mathbf{k})$~\eqref{green2} as
\begin{equation}\label{green3}
  G^{-1}(x, \mathbf{k}) = -\alpha \left[ x+ k^2 + t(x) \right].
\end{equation}
In the non-resonant regime $|b| \gg c^{1/3}$, when Eq.~\eqref{tb1} is applicable, the on-shell (Lorentzian) approximation yields $\gamma_\mathbf{k} = (\alpha \tilde{c}/b^2) k$  for quasiparticles damping which is equivalent to well-known result~\eqref{encor2} (see also Ref.~\cite{utesov2014}). Going beyond Lorentzian line-shape and approximate Eq.~\eqref{tb1} one can observe an asymmetry of the spectral weight and negligible $\propto \tilde{c}^2$ constant contribution to the broadening (see Fig.~\ref{fig4}). The former is related to significant self-energy $t(x)$ frequency variation near the quasiparticle pole (see, e.g., Ref.~\cite{toperverg1993}).

However, rather simple behaviour described above drastically changes in the resonant regime. To illustrate this analytically in the most simple way we consider $b=0$ case in the on-shell approximation. At $k \ll \tilde{c}^{1/3}$ the self-energy reads
\begin{equation}\label{green4}
  t(-k^2) \approx - e^{- i \pi/3} \tilde{c}^{2/3} + O(k^2).
\end{equation}
This means that the long-wavelength triplons damping is almost constant $\gamma_\mathbf{k} \approx \alpha (\sqrt{3}/2) \tilde{c}^{2/3}$ and these excitations are ill-defined. Moreover, it can be shown using Ioffe-Regel criterion~\cite{ioffe1960} that the modes with $k \ll c^{1/3}$ are localized. Indeed, mean free path reads $l_\mathbf{k} = v_\mathbf{k} \tau_\mathbf{k}  \propto k/c^{2/3}$ thus the condition of localization $ l_\mathbf{k} \lesssim \lambda_\mathbf{k} \propto 1/k$ is well satisfied. Notice, that $k \ll c^{1/3}$ is equivalent to the condition of quasiparticle wavelength to be much larger than the average distance between the impurities.
 
When impurities strength $u$ or equivalently parameter $b$ varies, one can once again observe the crossover between resonant and non-resonant regimes at $|b| \sim \tilde{c}^{1/3}$. In the crossover region linewidths are smaller than for $b=0$, but still well pronounced even for the mode with $k=0$.  We also point out that visible lineshape asymmetry for long-wavelength excitations can be observed in all regimes (see Fig.~\ref{fig4}) since the system DOS has singularities in the discussed frequency domain.

We illustrate the main results of this Subsec in Fig.~\ref{fig4}. We provide spectral weights \footnote{They are given by $\mathrm{Im} \, \Gamma(\omega, \mathbf{k})/\pi$ due to our definition of Green's function~\eqref{green1} with $-i 0 $ in the denominator.} in different regimes in Fig.~\ref{fig4} for $\tilde{c} = 0.01$ and $\alpha = 1$. One can see there pronounced asymmetry of the curves at small $k$, finite linewidths in resonant and crossover regimes, and conventional behaviour in the non-resonant case.

\section{Application to $\textrm{Ba}_{\textbf{3-x}}\textrm{Sr}_\textbf{x}\textrm{Cr}_\textbf{2}\textrm{O}_\textbf{8}$ }
\label{SAppl}

We adapt the model above to describe properties of Ba$_{3-x}$Sr$_x$Cr$_2$O$_8$ compound with small concentration of barium. In pure Sr$_3$Cr$_2$O$_8$ $J_0 \approx 5.55~\text{meV}$ and the shortest distance between magnetic Cr ions is $3.76 \AA$, whereas in Ba$_3$Cr$_2$O$_8$ $J_0 \approx 2.38~\text{meV}$ and the distance is $3.96 \AA$~\cite{grundmann2013}. Evidently barium ions makes the distance between magnetic ions larger. So, it is quite natural that small amount of barium ions result in weakening of neighboring intradimer couplings.

In order to describe experimentally observed gap renormalization in Ba$_{0.1}$Sr$_{2.9}$Cr$_2$O$_8$~\cite{gazizulina2017} we assume that in average there are $c = 1/30 \times 3 = 0.1$ unit cells with $u=(2.38-5.55)/3 \approx -1.06~\text{meV}$. Small deviations in interdimer couplings are neglected; we borrow the following parameters from Ref.~\cite{quintero2010} (all values are in meV): $J^\prime_1 = -0.04$, $J^{\prime\prime}_1 = 0.24$, $J^{\prime \prime \prime}_1 = 0.25$, $J^\prime_2 = 0.75$, $J^{\prime\prime}_2 = - 0.54$, $J^{\prime \prime \prime}_2 = -0.12$. Corresponding equation for $J(\mathbf{k})$ reads
\begin{eqnarray} \nonumber
   J_\mathbf{k} &=& J^\prime_1 \cos{k_c}+J^{\prime\prime}_1\cos{(k_c-k_a)}+J^{\prime \prime \prime}_1\cos{(k_c-k_a-k_b)}\\ && + J^\prime_2 \cos{k_a}+J^{\prime\prime}_2 \cos{k_b}+J^{\prime \prime \prime}_2 \cos{(k_a+k_b)}, \label{Jk}
\end{eqnarray}
where momenta components are written for monoclinic elementary cell with one dimer. Below we use simple RPA spectrum~\eqref{spec1} neglecting triplon-triplon interaction which can be treated using perturbation theory (see, e.g., Ref.~\cite{utesov2014jmmm}).

The theory of Sec.~\ref{STheor} is not directly applicable in this case because the spectrum is highly anisotropic near its minimum $\mathbf{k_0}$, $\varepsilon_\mathbf{k} \approx \Delta_0 + A_1 k^2_1 + A_2 k^2_2 + A_3 k^2_3$, $\Delta_0 = 3.451~\text{meV} $, $A_1=3.41499~\text{meV}$, $A_2=0.740279~\text{meV}$, $A_3=0.31128~\text{meV}$. Nevertheless, we can express the equation of SCTMA in the familiar form~\eqref{ST2} and, consequently,~\eqref{STeq1}, if the following observation is used:
\begin{eqnarray} \nonumber
  && \int \frac{d^3 q}{(2 \pi)^3} \left(\varepsilon_\mathbf{k} - \Delta_0 + [T(\omega) - \omega + \Delta_0] \right)^{-1} \\ && \approx C_1 - C_2 \sqrt{T(\omega) - \omega + \Delta_0} \label{Appl1}
\end{eqnarray}
with high accuracy if the imaginary part of $\sqrt{T(\omega) - \omega + \Delta_0}$ is negligible (c.f. Eq.~\eqref{int1}), $C_1 \approx 0.67$ and $C_2 \approx 0.3$. So, Eq.~\eqref{Appl1} is applicable for analysis of the gap renormalization, and the theory presented above can be used after we write SCTMA equation in the form
\begin{equation}\label{Appl2}
  T(\omega)=\frac{\tilde{c}}{b -\sqrt{T(\omega) + x }},
\end{equation}
with $\tilde{c} = c/C_2 \approx 0.34$ and $b = (1/u + C_1)/C_2 \approx - 0.94$. The last equality shows that we are approximately in the regime of $|b| \approx 3 \tilde{c}^{1/3}/2$ without bound states inside the gap. Then, the solution for $T$ is given by Eq.~\eqref{tb1}. One obtains in this case
\begin{equation}\label{Appl3}
  \Delta = \Delta_0 + \frac{\tilde{c}}{b} \approx 3.1~\text{K}.
\end{equation}
We notice, that exact solution of cubic equation (see Appendix) yields close value $\Delta \approx 3.125~\text{K}$.

Both results for gap renormalization lies in a good agreement with experimental observation $\Delta \approx 3.174~\text{meV}$ which was made using inelastic neutron scattering in Ref.~\cite{gazizulina2017}.

\section{Summary}
\label{SSum}

To conclude, we developed a theory describing magnetic excitations in gapped phases of quantum magnets with bond disorder characterized by parameter $u$ (e.g., the difference between intradimer coupling on the defect bond and on the regular one in spin-dimer compounds). Using self-consistent T-matrix approximation we discuss quasiparticles gap renormalization and (if exist) broadening of the localized on impurity states.

For positive or small negative $u$ the results are rather trivial. The correction to the gap is proportional to impurities concentration $c$ (see Eq.~\eqref{gapbB}) and has the same sign with $u$. However, when $u$ tends to the threshold value for bound on impurities states appearance the situation drastically changes. This regime is governed by the resonant scattering off impurities which yields gap renormalization $\propto c^{2/3}$~\eqref{gapb0}. Under further $u$ decreasing beyond the threshold value in the resonant scattering regime system DOS is highly nontrivial, having the form of overlapped band DOS and semi-circle for localized on impurities states (see Fig.~\ref{fig3}). Finally, at large negative $u$ semi-circle for isolated impurity levels is well separated from the band DOS which, however, is characterized by positive correction to the gap~\eqref{gapbB} due to some sort of repulsion between this two parts of the system DOS. Furthermore, our theory successfully describes the boundary between the latter two regimes.

Long-wavelength elementary excitation were shown to be ill-defined in the resonant regime and the crossover region. Their damping is almost constant, $\gamma_{\mathbf{k}} \propto c^{2/3} + O(k^2)$. Using Ioffe-Regel criterion we show that excitations with $k \ll c^{1/3}$ are localized. In the non-resonant regime we justify previous results of Ref.~\cite{utesov2014}, $\gamma_{\mathbf{k}} \propto c k$. In all the regimes visible asymmetry of the long-wavelength excitations spectral weights is reported, which is related to significant variations in the self-energy near the DOS singulartities.

We show the applicability of the present theory for gapped phases of quantum magnets by successful quantitative description of the gap renormalization in Ba$_{3-x}$Sr$_x$Cr$_2$O$_8$ at $x=2.9$ which was observed experimentally in recent paper~\cite{gazizulina2017}. The developed theory can be also applied for analysis of experimental data on other types of quantum magnets with bond disorder.

\begin{acknowledgments}

We are grateful to A.~V.~Syromyatnikov and A.~G.~Yashenkin for valuable discussions. The reported study was supported by the Foundation for the Advancement of Theoretical Physics and Mathematics ``BASIS''.

\end{acknowledgments}

\appendix

\section{Solution of SCTMA equation}
\label{SAppend}

Here we present exact solutions of SCTMA equation~\eqref{STeq1} using Cardano's formula. All the discussed in Sec.~\ref{STheor} results can be obtained using the formulas below. However, one should be accurate and check whether considered solution at given $x$ is physical or spurious. For physical solution the following equality should hold in the corresponding frequency region:
\begin{equation}\label{Acond1}
  \sqrt{t+x} = b-\frac{\tilde{c}}{t}.
\end{equation}
In order to make formulas for Eq.~\eqref{STeq1} solutions simpler we introduce
\begin{eqnarray}
  \tilde{x} &=& x - b^2, \\
  \beta &=& \tilde{x}^2 - 6 b \tilde{c}, \\
  \gamma &=& 27 \tilde{c}^2 + 18 b \tilde{c} \tilde{x} - 2 \tilde{x}^3.
\end{eqnarray}
Then, all three solutions of Eq.~\eqref{STeq1} can be written as
\begin{equation}\label{Asol1}
  t = - \frac{\tilde{x}}{3} - \frac{(-1)^{1/3} 2^{1/3} \beta}{3 \left(\gamma+ \sqrt{\gamma^2 - 4 \beta^3}\right)^{1/3}} - \frac{ \left(\gamma + \sqrt{\gamma^2 - 4 \beta^3}\right)^{1/3}}{(-1)^{1/3} 2^{1/3} 3 },
\end{equation}
where $(-1)^{1/3} = -1, (1 + i\sqrt{3})/2, (1-i\sqrt{3})/2$, each value should be taken simultaneously in both second and third term of this equation. Other cube roots should be taken for the branch $(-1)^{1/3}=-1$.

Next, it can be shown that points dividing zero and nonzero DOS regions (one of them corresponds to the renormalized band gap) satisfy the condition
\begin{equation}\label{Acond2}
  \gamma^2 = 4 \beta^3.
\end{equation}
It can be also solved as a cubic one after some transformations. Denoting
\begin{equation}
  \xi = 8 b^6 - 540 b^3 \tilde{c} - 729 \tilde{c}^2 + \sqrt{27 \tilde{c}(27\tilde{c} - 8b ^3)^3},
\end{equation}
one can write solutions for $x$ as follows:
\begin{equation}\label{Asol2}
  x = \frac{2 b^2}{3} - \frac{(-1)^{1/3} 2 b (b^3 + 27 b \tilde{c})}{3 \xi^{1/3}} - \frac{ \xi^{1/3}}{ (-1)^{1/3} 6 },
\end{equation}
notation is the same with Eq.~\eqref{Asol1}. From this solution important curve in $(\tilde{c},b)$-parameters space arises, $b = 3 \tilde{c}^{1/3}/2$. Mathematically, it divides regions with one and three real solutions for~\eqref{Asol2}. Physically it determines either DOS of the band and localized on impurities states are overlapped (one real solution, $b < 3 \tilde{c}^{1/3}/2$) or separated from each other (three real solutions, $b > 3 \tilde{c}^{1/3}/2$).

\bibliography{References}

\begin{thebibliography}{36}
\expandafter\ifx\csname natexlab\endcsname\relax\def\natexlab#1{#1}\fi
\expandafter\ifx\csname bibnamefont\endcsname\relax
  \def\bibnamefont#1{#1}\fi
\expandafter\ifx\csname bibfnamefont\endcsname\relax
  \def\bibfnamefont#1{#1}\fi
\expandafter\ifx\csname citenamefont\endcsname\relax
  \def\citenamefont#1{#1}\fi
\expandafter\ifx\csname url\endcsname\relax
  \def\url#1{\texttt{#1}}\fi
\expandafter\ifx\csname urlprefix\endcsname\relax\def\urlprefix{URL }\fi
\providecommand{\bibinfo}[2]{#2}
\providecommand{\eprint}[2][]{\url{#2}}

\bibitem[{\citenamefont{Vasiliev et~al.}(2018)\citenamefont{Vasiliev, Volkova,
  Zvereva, and Markina}}]{vasiliev2018}
\bibinfo{author}{\bibfnamefont{A.}~\bibnamefont{Vasiliev}},
  \bibinfo{author}{\bibfnamefont{O.}~\bibnamefont{Volkova}},
  \bibinfo{author}{\bibfnamefont{E.}~\bibnamefont{Zvereva}}, \bibnamefont{and}
  \bibinfo{author}{\bibfnamefont{M.}~\bibnamefont{Markina}},
  \bibinfo{journal}{npj Quantum Materials} \textbf{\bibinfo{volume}{3}},
  \bibinfo{pages}{1} (\bibinfo{year}{2018}).

\bibitem[{\citenamefont{Sachdev}(2011)}]{sachdev2011}
\bibinfo{author}{\bibfnamefont{S.}~\bibnamefont{Sachdev}},
  \emph{\bibinfo{title}{Quantum Phase Transitions}}
  (\bibinfo{publisher}{Cambridge University Press}, \bibinfo{year}{2011}),
  \bibinfo{edition}{2nd} ed.

\bibitem[{\citenamefont{Giamarchi et~al.}(2008)\citenamefont{Giamarchi,
  R{\"u}egg, and Tchernyshyov}}]{giamarchi2008bose}
\bibinfo{author}{\bibfnamefont{T.}~\bibnamefont{Giamarchi}},
  \bibinfo{author}{\bibfnamefont{C.}~\bibnamefont{R{\"u}egg}},
  \bibnamefont{and}
  \bibinfo{author}{\bibfnamefont{O.}~\bibnamefont{Tchernyshyov}},
  \bibinfo{journal}{Nature Physics} \textbf{\bibinfo{volume}{4}},
  \bibinfo{pages}{198} (\bibinfo{year}{2008}).

\bibitem[{\citenamefont{Zheludev and Roscilde}(2013)}]{zheludev2013dirty}
\bibinfo{author}{\bibfnamefont{A.}~\bibnamefont{Zheludev}} \bibnamefont{and}
  \bibinfo{author}{\bibfnamefont{T.}~\bibnamefont{Roscilde}},
  \bibinfo{journal}{Comptes Rendus Physique} \textbf{\bibinfo{volume}{14}},
  \bibinfo{pages}{740} (\bibinfo{year}{2013}).

\bibitem[{\citenamefont{Fisher et~al.}(1989)\citenamefont{Fisher, Weichman,
  Grinstein, and Fisher}}]{fisher1989}
\bibinfo{author}{\bibfnamefont{M.~P.} \bibnamefont{Fisher}},
  \bibinfo{author}{\bibfnamefont{P.~B.} \bibnamefont{Weichman}},
  \bibinfo{author}{\bibfnamefont{G.}~\bibnamefont{Grinstein}},
  \bibnamefont{and} \bibinfo{author}{\bibfnamefont{D.~S.}
  \bibnamefont{Fisher}}, \bibinfo{journal}{Physical Review B}
  \textbf{\bibinfo{volume}{40}}, \bibinfo{pages}{546} (\bibinfo{year}{1989}).

\bibitem[{\citenamefont{Yu et~al.}(2012)\citenamefont{Yu, Yin, Sullivan, Xia,
  Huan, Paduan-Filho, Oliveira~Jr, Haas, Steppke, Miclea et~al.}}]{yu2012bose}
\bibinfo{author}{\bibfnamefont{R.}~\bibnamefont{Yu}},
  \bibinfo{author}{\bibfnamefont{L.}~\bibnamefont{Yin}},
  \bibinfo{author}{\bibfnamefont{N.~S.} \bibnamefont{Sullivan}},
  \bibinfo{author}{\bibfnamefont{J.}~\bibnamefont{Xia}},
  \bibinfo{author}{\bibfnamefont{C.}~\bibnamefont{Huan}},
  \bibinfo{author}{\bibfnamefont{A.}~\bibnamefont{Paduan-Filho}},
  \bibinfo{author}{\bibfnamefont{N.~F.} \bibnamefont{Oliveira~Jr}},
  \bibinfo{author}{\bibfnamefont{S.}~\bibnamefont{Haas}},
  \bibinfo{author}{\bibfnamefont{A.}~\bibnamefont{Steppke}},
  \bibinfo{author}{\bibfnamefont{C.~F.} \bibnamefont{Miclea}},
  \bibnamefont{et~al.}, \bibinfo{journal}{Nature}
  \textbf{\bibinfo{volume}{489}}, \bibinfo{pages}{379} (\bibinfo{year}{2012}).

\bibitem[{\citenamefont{Pollet et~al.}(2009)\citenamefont{Pollet, Prokof'ev,
  Svistunov, and Troyer}}]{theorem}
\bibinfo{author}{\bibfnamefont{L.}~\bibnamefont{Pollet}},
  \bibinfo{author}{\bibfnamefont{N.~V.} \bibnamefont{Prokof'ev}},
  \bibinfo{author}{\bibfnamefont{B.~V.} \bibnamefont{Svistunov}},
  \bibnamefont{and} \bibinfo{author}{\bibfnamefont{M.}~\bibnamefont{Troyer}},
  \bibinfo{journal}{Phys. Rev. Lett.} \textbf{\bibinfo{volume}{103}},
  \bibinfo{pages}{140402} (\bibinfo{year}{2009}).

\bibitem[{\citenamefont{Kr\"uger et~al.}(2011)\citenamefont{Kr\"uger, Hong, and
  Phillips}}]{kruger2011}
\bibinfo{author}{\bibfnamefont{F.}~\bibnamefont{Kr\"uger}},
  \bibinfo{author}{\bibfnamefont{S.}~\bibnamefont{Hong}}, \bibnamefont{and}
  \bibinfo{author}{\bibfnamefont{P.}~\bibnamefont{Phillips}},
  \bibinfo{journal}{Phys. Rev. B} \textbf{\bibinfo{volume}{84}},
  \bibinfo{pages}{115118} (\bibinfo{year}{2011}),
  \urlprefix\url{https://link.aps.org/doi/10.1103/PhysRevB.84.115118}.

\bibitem[{\citenamefont{Thomson and Kr\"uger}(2015)}]{thomson2015}
\bibinfo{author}{\bibfnamefont{S.~J.} \bibnamefont{Thomson}} \bibnamefont{and}
  \bibinfo{author}{\bibfnamefont{F.}~\bibnamefont{Kr\"uger}},
  \bibinfo{journal}{Phys. Rev. B} \textbf{\bibinfo{volume}{92}},
  \bibinfo{pages}{180201} (\bibinfo{year}{2015}),
  \urlprefix\url{https://link.aps.org/doi/10.1103/PhysRevB.92.180201}.

\bibitem[{\citenamefont{Syromyatnikov and Sizanov}(2017)}]{syromyatnikov2017}
\bibinfo{author}{\bibfnamefont{A.}~\bibnamefont{Syromyatnikov}}
  \bibnamefont{and} \bibinfo{author}{\bibfnamefont{A.}~\bibnamefont{Sizanov}},
  \bibinfo{journal}{Physical Review B} \textbf{\bibinfo{volume}{95}},
  \bibinfo{pages}{014206} (\bibinfo{year}{2017}).

\bibitem[{\citenamefont{Syromyatnikov}(2017)}]{syromyatnikov2017Mott}
\bibinfo{author}{\bibfnamefont{A.}~\bibnamefont{Syromyatnikov}},
  \bibinfo{journal}{Annalen der Physik} \textbf{\bibinfo{volume}{529}},
  \bibinfo{pages}{1700055} (\bibinfo{year}{2017}).

\bibitem[{\citenamefont{Yashenkin et~al.}(2016)\citenamefont{Yashenkin, Utesov,
  Sizanov, and Syromyatnikov}}]{yashenkin2016}
\bibinfo{author}{\bibfnamefont{A.}~\bibnamefont{Yashenkin}},
  \bibinfo{author}{\bibfnamefont{O.}~\bibnamefont{Utesov}},
  \bibinfo{author}{\bibfnamefont{A.}~\bibnamefont{Sizanov}}, \bibnamefont{and}
  \bibinfo{author}{\bibfnamefont{A.}~\bibnamefont{Syromyatnikov}},
  \bibinfo{journal}{Journal of Magnetism and Magnetic Materials}
  \textbf{\bibinfo{volume}{397}}, \bibinfo{pages}{11 } (\bibinfo{year}{2016}),
  ISSN \bibinfo{issn}{0304-8853},
  \urlprefix\url{http://www.sciencedirect.com/science/article/pii/S0304885315304923}.

\bibitem[{\citenamefont{Hong et~al.}(2010)\citenamefont{Hong, Zheludev, Manaka,
  and Regnault}}]{hong2010}
\bibinfo{author}{\bibfnamefont{T.}~\bibnamefont{Hong}},
  \bibinfo{author}{\bibfnamefont{A.}~\bibnamefont{Zheludev}},
  \bibinfo{author}{\bibfnamefont{H.}~\bibnamefont{Manaka}}, \bibnamefont{and}
  \bibinfo{author}{\bibfnamefont{L.-P.} \bibnamefont{Regnault}},
  \bibinfo{journal}{Phys. Rev. B} \textbf{\bibinfo{volume}{81}},
  \bibinfo{pages}{060410} (\bibinfo{year}{2010}),
  \urlprefix\url{https://link.aps.org/doi/10.1103/PhysRevB.81.060410}.

\bibitem[{\citenamefont{H\"uvonen et~al.}(2012)\citenamefont{H\"uvonen, Zhao,
  M\aa{}nsson, Yankova, Ressouche, Niedermayer, Laver, Gvasaliya, and
  Zheludev}}]{huvonen2012}
\bibinfo{author}{\bibfnamefont{D.}~\bibnamefont{H\"uvonen}},
  \bibinfo{author}{\bibfnamefont{S.}~\bibnamefont{Zhao}},
  \bibinfo{author}{\bibfnamefont{M.}~\bibnamefont{M\aa{}nsson}},
  \bibinfo{author}{\bibfnamefont{T.}~\bibnamefont{Yankova}},
  \bibinfo{author}{\bibfnamefont{E.}~\bibnamefont{Ressouche}},
  \bibinfo{author}{\bibfnamefont{C.}~\bibnamefont{Niedermayer}},
  \bibinfo{author}{\bibfnamefont{M.}~\bibnamefont{Laver}},
  \bibinfo{author}{\bibfnamefont{S.~N.} \bibnamefont{Gvasaliya}},
  \bibnamefont{and} \bibinfo{author}{\bibfnamefont{A.}~\bibnamefont{Zheludev}},
  \bibinfo{journal}{Phys. Rev. B} \textbf{\bibinfo{volume}{85}},
  \bibinfo{pages}{100410} (\bibinfo{year}{2012}),
  \urlprefix\url{https://link.aps.org/doi/10.1103/PhysRevB.85.100410}.

\bibitem[{\citenamefont{Grundmann et~al.}(2013)\citenamefont{Grundmann,
  Schilling, Marjerrison, Dabkowska, and Gaulin}}]{grundmann2013}
\bibinfo{author}{\bibfnamefont{H.}~\bibnamefont{Grundmann}},
  \bibinfo{author}{\bibfnamefont{A.}~\bibnamefont{Schilling}},
  \bibinfo{author}{\bibfnamefont{C.~A.} \bibnamefont{Marjerrison}},
  \bibinfo{author}{\bibfnamefont{H.~A.} \bibnamefont{Dabkowska}},
  \bibnamefont{and} \bibinfo{author}{\bibfnamefont{B.~D.}
  \bibnamefont{Gaulin}}, \bibinfo{journal}{Materials Research Bulletin}
  \textbf{\bibinfo{volume}{48}}, \bibinfo{pages}{3108 } (\bibinfo{year}{2013}),
  ISSN \bibinfo{issn}{0025-5408},
  \urlprefix\url{http://www.sciencedirect.com/science/article/pii/S002554081300353X}.

\bibitem[{\citenamefont{Povarov et~al.}(2015)\citenamefont{Povarov, Wulf,
  H{\"u}vonen, Ollivier, Paduan-Filho, and Zheludev}}]{povarov2015dynamics}
\bibinfo{author}{\bibfnamefont{K.~Y.} \bibnamefont{Povarov}},
  \bibinfo{author}{\bibfnamefont{E.}~\bibnamefont{Wulf}},
  \bibinfo{author}{\bibfnamefont{D.}~\bibnamefont{H{\"u}vonen}},
  \bibinfo{author}{\bibfnamefont{J.}~\bibnamefont{Ollivier}},
  \bibinfo{author}{\bibfnamefont{A.}~\bibnamefont{Paduan-Filho}},
  \bibnamefont{and} \bibinfo{author}{\bibfnamefont{A.}~\bibnamefont{Zheludev}},
  \bibinfo{journal}{Physical Review B} \textbf{\bibinfo{volume}{92}},
  \bibinfo{pages}{024429} (\bibinfo{year}{2015}).

\bibitem[{\citenamefont{Povarov et~al.}(2017)\citenamefont{Povarov, Mannig,
  Perren, M\"oller, Wulf, Ollivier, and Zheludev}}]{Povarov2017}
\bibinfo{author}{\bibfnamefont{K.~Y.} \bibnamefont{Povarov}},
  \bibinfo{author}{\bibfnamefont{A.}~\bibnamefont{Mannig}},
  \bibinfo{author}{\bibfnamefont{G.}~\bibnamefont{Perren}},
  \bibinfo{author}{\bibfnamefont{J.~S.} \bibnamefont{M\"oller}},
  \bibinfo{author}{\bibfnamefont{E.}~\bibnamefont{Wulf}},
  \bibinfo{author}{\bibfnamefont{J.}~\bibnamefont{Ollivier}}, \bibnamefont{and}
  \bibinfo{author}{\bibfnamefont{A.}~\bibnamefont{Zheludev}},
  \bibinfo{journal}{Phys. Rev. B} \textbf{\bibinfo{volume}{96}},
  \bibinfo{pages}{140414} (\bibinfo{year}{2017}),
  \urlprefix\url{https://link.aps.org/doi/10.1103/PhysRevB.96.140414}.

\bibitem[{\citenamefont{Utesov et~al.}(2014)\citenamefont{Utesov, Sizanov, and
  Syromyatnikov}}]{utesov2014}
\bibinfo{author}{\bibfnamefont{O.~I.} \bibnamefont{Utesov}},
  \bibinfo{author}{\bibfnamefont{A.~V.} \bibnamefont{Sizanov}},
  \bibnamefont{and} \bibinfo{author}{\bibfnamefont{A.~V.}
  \bibnamefont{Syromyatnikov}}, \bibinfo{journal}{Phys. Rev. B}
  \textbf{\bibinfo{volume}{90}}, \bibinfo{pages}{155121}
  (\bibinfo{year}{2014}),
  \urlprefix\url{https://link.aps.org/doi/10.1103/PhysRevB.90.155121}.

\bibitem[{\citenamefont{Izyumov and Medvedev}(1973)}]{izyumov}
\bibinfo{author}{\bibfnamefont{Y.~A.} \bibnamefont{Izyumov}} \bibnamefont{and}
  \bibinfo{author}{\bibfnamefont{M.}~\bibnamefont{Medvedev}},
  \emph{\bibinfo{title}{Magnetically Ordered Crystals Containing Impurities}}
  (\bibinfo{publisher}{Consultants Bureau}, \bibinfo{address}{New York},
  \bibinfo{year}{1973}).

\bibitem[{\citenamefont{Doniach and Sondheimer}(1998)}]{doniach1998}
\bibinfo{author}{\bibfnamefont{S.}~\bibnamefont{Doniach}} \bibnamefont{and}
  \bibinfo{author}{\bibfnamefont{E.}~\bibnamefont{Sondheimer}},
  \emph{\bibinfo{title}{Green's Functions for Solid State Physicists}}
  (\bibinfo{publisher}{Imperial College Press}, \bibinfo{year}{1998}), ISBN
  \bibinfo{isbn}{9781860940804},
  \urlprefix\url{https://books.google.ru/books?id=8OmM-\_pQJgAC}.

\bibitem[{\citenamefont{Lee}(1993)}]{lee1993}
\bibinfo{author}{\bibfnamefont{P.~A.} \bibnamefont{Lee}},
  \bibinfo{journal}{Physical review letters} \textbf{\bibinfo{volume}{71}},
  \bibinfo{pages}{1887} (\bibinfo{year}{1993}).

\bibitem[{\citenamefont{Ostrovsky et~al.}(2006)\citenamefont{Ostrovsky, Gornyi,
  and Mirlin}}]{ostrovsky2006}
\bibinfo{author}{\bibfnamefont{P.}~\bibnamefont{Ostrovsky}},
  \bibinfo{author}{\bibfnamefont{I.}~\bibnamefont{Gornyi}}, \bibnamefont{and}
  \bibinfo{author}{\bibfnamefont{A.}~\bibnamefont{Mirlin}},
  \bibinfo{journal}{Physical Review B} \textbf{\bibinfo{volume}{74}},
  \bibinfo{pages}{235443} (\bibinfo{year}{2006}).

\bibitem[{\citenamefont{Yashenkin et~al.}(2001)\citenamefont{Yashenkin,
  Atkinson, Gornyi, Hirschfeld, and Khveshchenko}}]{yashenkin2001}
\bibinfo{author}{\bibfnamefont{A.~G.} \bibnamefont{Yashenkin}},
  \bibinfo{author}{\bibfnamefont{W.~A.} \bibnamefont{Atkinson}},
  \bibinfo{author}{\bibfnamefont{I.~V.} \bibnamefont{Gornyi}},
  \bibinfo{author}{\bibfnamefont{P.~J.} \bibnamefont{Hirschfeld}},
  \bibnamefont{and} \bibinfo{author}{\bibfnamefont{D.~V.}
  \bibnamefont{Khveshchenko}}, \bibinfo{journal}{Phys. Rev. Lett.}
  \textbf{\bibinfo{volume}{86}}, \bibinfo{pages}{5982} (\bibinfo{year}{2001}),
  \urlprefix\url{https://link.aps.org/doi/10.1103/PhysRevLett.86.5982}.

\bibitem[{\citenamefont{Ando and Uemura}(1974)}]{ando1974}
\bibinfo{author}{\bibfnamefont{T.}~\bibnamefont{Ando}} \bibnamefont{and}
  \bibinfo{author}{\bibfnamefont{Y.}~\bibnamefont{Uemura}},
  \bibinfo{journal}{Journal of the Physical Society of Japan}
  \textbf{\bibinfo{volume}{36}}, \bibinfo{pages}{959} (\bibinfo{year}{1974}).

\bibitem[{\citenamefont{Gazizulina et~al.}(2017)\citenamefont{Gazizulina,
  Quintero-Castro, and Schilling}}]{gazizulina2017}
\bibinfo{author}{\bibfnamefont{A.}~\bibnamefont{Gazizulina}},
  \bibinfo{author}{\bibfnamefont{D.~L.} \bibnamefont{Quintero-Castro}},
  \bibnamefont{and}
  \bibinfo{author}{\bibfnamefont{A.}~\bibnamefont{Schilling}},
  \bibinfo{journal}{Phys. Rev. B} \textbf{\bibinfo{volume}{96}},
  \bibinfo{pages}{184201} (\bibinfo{year}{2017}),
  \urlprefix\url{https://link.aps.org/doi/10.1103/PhysRevB.96.184201}.

\bibitem[{\citenamefont{Utesov and Syromyatnikov}(2014)}]{utesov2014jmmm}
\bibinfo{author}{\bibfnamefont{O.}~\bibnamefont{Utesov}} \bibnamefont{and}
  \bibinfo{author}{\bibfnamefont{A.}~\bibnamefont{Syromyatnikov}},
  \bibinfo{journal}{Journal of Magnetism and Magnetic Materials}
  \textbf{\bibinfo{volume}{358-359}}, \bibinfo{pages}{177 }
  (\bibinfo{year}{2014}), ISSN \bibinfo{issn}{0304-8853},
  \urlprefix\url{http://www.sciencedirect.com/science/article/pii/S0304885314000560}.

\bibitem[{\citenamefont{Sachdev and Bhatt}(1990)}]{sachdev1990}
\bibinfo{author}{\bibfnamefont{S.}~\bibnamefont{Sachdev}} \bibnamefont{and}
  \bibinfo{author}{\bibfnamefont{R.~N.} \bibnamefont{Bhatt}},
  \bibinfo{journal}{Physical Review B} \textbf{\bibinfo{volume}{41}},
  \bibinfo{pages}{9323} (\bibinfo{year}{1990}).

\bibitem[{\citenamefont{Kotov et~al.}(1998)\citenamefont{Kotov, Sushkov,
  Weihong, and Oitmaa}}]{kotov1998}
\bibinfo{author}{\bibfnamefont{V.}~\bibnamefont{Kotov}},
  \bibinfo{author}{\bibfnamefont{O.}~\bibnamefont{Sushkov}},
  \bibinfo{author}{\bibfnamefont{Z.}~\bibnamefont{Weihong}}, \bibnamefont{and}
  \bibinfo{author}{\bibfnamefont{J.}~\bibnamefont{Oitmaa}},
  \bibinfo{journal}{Physical review letters} \textbf{\bibinfo{volume}{80}},
  \bibinfo{pages}{5790} (\bibinfo{year}{1998}).

\bibitem[{\citenamefont{Kofu et~al.}(2009)\citenamefont{Kofu, Ueda, Nojiri,
  Oshima, Zenmoto, Rule, Gerischer, Lake, Batista, Ueda et~al.}}]{kofu2009}
\bibinfo{author}{\bibfnamefont{M.}~\bibnamefont{Kofu}},
  \bibinfo{author}{\bibfnamefont{H.}~\bibnamefont{Ueda}},
  \bibinfo{author}{\bibfnamefont{H.}~\bibnamefont{Nojiri}},
  \bibinfo{author}{\bibfnamefont{Y.}~\bibnamefont{Oshima}},
  \bibinfo{author}{\bibfnamefont{T.}~\bibnamefont{Zenmoto}},
  \bibinfo{author}{\bibfnamefont{K.~C.} \bibnamefont{Rule}},
  \bibinfo{author}{\bibfnamefont{S.}~\bibnamefont{Gerischer}},
  \bibinfo{author}{\bibfnamefont{B.}~\bibnamefont{Lake}},
  \bibinfo{author}{\bibfnamefont{C.~D.} \bibnamefont{Batista}},
  \bibinfo{author}{\bibfnamefont{Y.}~\bibnamefont{Ueda}}, \bibnamefont{et~al.},
  \bibinfo{journal}{Phys. Rev. Lett.} \textbf{\bibinfo{volume}{102}},
  \bibinfo{pages}{177204} (\bibinfo{year}{2009}),
  \urlprefix\url{https://link.aps.org/doi/10.1103/PhysRevLett.102.177204}.

\bibitem[{\citenamefont{Zittartz and Langer}(1966)}]{zittartz1966}
\bibinfo{author}{\bibfnamefont{J.}~\bibnamefont{Zittartz}} \bibnamefont{and}
  \bibinfo{author}{\bibfnamefont{J.~S.} \bibnamefont{Langer}},
  \bibinfo{journal}{Phys. Rev.} \textbf{\bibinfo{volume}{148}},
  \bibinfo{pages}{741} (\bibinfo{year}{1966}),
  \urlprefix\url{https://link.aps.org/doi/10.1103/PhysRev.148.741}.

\bibitem[{\citenamefont{Halperin and Lax}(1966)}]{halperin1966}
\bibinfo{author}{\bibfnamefont{B.~I.} \bibnamefont{Halperin}} \bibnamefont{and}
  \bibinfo{author}{\bibfnamefont{M.}~\bibnamefont{Lax}},
  \bibinfo{journal}{Phys. Rev.} \textbf{\bibinfo{volume}{148}},
  \bibinfo{pages}{722} (\bibinfo{year}{1966}),
  \urlprefix\url{https://link.aps.org/doi/10.1103/PhysRev.148.722}.

\bibitem[{\citenamefont{Halperin and Lax}(1967)}]{halperin1967}
\bibinfo{author}{\bibfnamefont{B.~I.} \bibnamefont{Halperin}} \bibnamefont{and}
  \bibinfo{author}{\bibfnamefont{M.}~\bibnamefont{Lax}},
  \bibinfo{journal}{Phys. Rev.} \textbf{\bibinfo{volume}{153}},
  \bibinfo{pages}{802} (\bibinfo{year}{1967}),
  \urlprefix\url{https://link.aps.org/doi/10.1103/PhysRev.153.802}.

\bibitem[{\citenamefont{Lifshitz}(1968)}]{lifshitz1968}
\bibinfo{author}{\bibfnamefont{I.}~\bibnamefont{Lifshitz}},
  \bibinfo{journal}{Sov. Phys. JETP} \textbf{\bibinfo{volume}{26}},
  \bibinfo{pages}{012110} (\bibinfo{year}{1968}).

\bibitem[{\citenamefont{Toperverg and Yashenkin}(1993)}]{toperverg1993}
\bibinfo{author}{\bibfnamefont{B.~P.} \bibnamefont{Toperverg}}
  \bibnamefont{and} \bibinfo{author}{\bibfnamefont{A.~G.}
  \bibnamefont{Yashenkin}}, \bibinfo{journal}{Phys. Rev. B}
  \textbf{\bibinfo{volume}{48}}, \bibinfo{pages}{16505} (\bibinfo{year}{1993}),
  \urlprefix\url{https://link.aps.org/doi/10.1103/PhysRevB.48.16505}.

\bibitem[{\citenamefont{Ioffe and Regel}(1960)}]{ioffe1960}
\bibinfo{author}{\bibfnamefont{A.}~\bibnamefont{Ioffe}} \bibnamefont{and}
  \bibinfo{author}{\bibfnamefont{A.}~\bibnamefont{Regel}},
  \emph{\bibinfo{title}{Progress in semiconductors vol. 4}}
  (\bibinfo{year}{1960}).

\bibitem[{\citenamefont{Quintero-Castro
  et~al.}(2010)\citenamefont{Quintero-Castro, Lake, Wheeler, Islam, Guidi,
  Rule, Izaola, Russina, Kiefer, and Skourski}}]{quintero2010}
\bibinfo{author}{\bibfnamefont{D.~L.} \bibnamefont{Quintero-Castro}},
  \bibinfo{author}{\bibfnamefont{B.}~\bibnamefont{Lake}},
  \bibinfo{author}{\bibfnamefont{E.~M.} \bibnamefont{Wheeler}},
  \bibinfo{author}{\bibfnamefont{A.~T. M.~N.} \bibnamefont{Islam}},
  \bibinfo{author}{\bibfnamefont{T.}~\bibnamefont{Guidi}},
  \bibinfo{author}{\bibfnamefont{K.~C.} \bibnamefont{Rule}},
  \bibinfo{author}{\bibfnamefont{Z.}~\bibnamefont{Izaola}},
  \bibinfo{author}{\bibfnamefont{M.}~\bibnamefont{Russina}},
  \bibinfo{author}{\bibfnamefont{K.}~\bibnamefont{Kiefer}}, \bibnamefont{and}
  \bibinfo{author}{\bibfnamefont{Y.}~\bibnamefont{Skourski}},
  \bibinfo{journal}{Phys. Rev. B} \textbf{\bibinfo{volume}{81}},
  \bibinfo{pages}{014415} (\bibinfo{year}{2010}),
  \urlprefix\url{https://link.aps.org/doi/10.1103/PhysRevB.81.014415}.

\end{thebibliography}

\end{document}